# Evidence for a White-light Flare on 10 September 1886

J.M. Vaquero[1,2] ● M. Vázquez[3,4] ● J. Sánchez Almeida[3,4]

[1] Departamento de Física, Universidad de Extremadura, Avda. Santa Teresa de Jornet 38, 06800 Mérida, Spain, jvaquero@unex.es

[2] Instituto Universitario de Investigación del Agua, Cambio Climático y Sostenibilidad (IACYS), Universidad de Extremadura, 06006 Badajoz, Spain

[3] Instituto de Astrofísica de Canarias, 38200 La Laguna, Tenerife, Spain

[4] Departamento de Astrofísica, Universidad de La Laguna, 38205 La Laguna, Spain

**Abstract.** We present evidence for the occurrence of a white-light flare on 10 September 1886. It represents the third of such rare events reported in the history of astronomy. The flare was mentioned by Valderrama (1886, *L'Astronomie* **5**, 388). In this article we have used the original logbook of the observer, J. Valderrama y Aguilar, an amateur astronomer that lived in Madrid and Santa Cruz de Tenerife at that time.

**Keywords.** Solar Flares; White-Light Flares

## 1. Introduction

Historical observations of the Sun can provide evidence for rare observational events (Vaquero and Vázquez, 2009). White-light flares (hereafter WLFs) have been historically defined as solar flares observed in integrated light against the background solar disk. They were presented as the most energetic flares, for which the released magnetic energy reaches the photosphere, and so they show up in the continuum. Due to the lack of dedicated observational programs of this phenomenon, WLF observations were very rare in the past. Neiding and Cliver (1983) provided the most complete list of recorded WLFs from 1859 to 1982, including only four observed WLFs in the 19th century. These occurred on 1 September 1859 (Carrington, 1859; Hogdson, 1859; Cliver and Dietrich, 2013), 13 November 1872 (Secchi, 1872), 17 June 1891 (Trouvelot, 1891), and 15 July 1892 (Rudaux, 1892). Several studies of WLFs observed with instruments onboard spacecraft have been recently done (Matthews *et al.*, 2003; Hudson, Wolfson, and Metcalf, 2006; Wang, 2009). Modern studies suggest that white-light emission is a common feature of all solar flares (Kuhar *et al.*, 2016), but high sensitivity is needed to detect it. For example, Jess *et al.* (2008) observed a white-light brightening associated with a C2.0 flare event. In any case, the search for WLFs has changed drastically with the use of spacecraft and it is not easy to compare modern and historical data.

The aim of this short article is to present evidence for a WLF observed by the Spanish amateur astronomer Juan Valderrama y Aguilar (hereafter Valderrama). The observations were made from Madrid on the 10 September 1886, using a modest 6.6 cm telescope.

The handwritten observing logbooks of Valderrama are preserved in the Library of the Instituto de Astrofísica de Canarias (IAC) in La Laguna (Spain) and correspond to a continuous period from December 1885 to April 1888, with some additional annotations later on. Only a few details are known about Valderrama and his astronomical observations (Oliver, 1997). He was born in Santa Cruz de Tenerife on 15 February 1869. His family moved to Madrid where they lived from 1882 until 1888, when his father died and the family returned to Santa Cruz de Tenerife. Valderrama carried out different meteorological observations, and was involved in the creation of the high-altitude Meteorological Observatory at Izaña (2400 m above the sea level), which is still in operation. He died in Santa Cruz on 17 March 1912.

**2. Description of the Phenomenon**

On 10 September 1886, Valderrama describes a "strange" observation using his modest 6.6 cm refractor telescope. See the transcription of the original manuscript, in Spanish, and the English translation in Appendix 1 and 2 of this article, respectively. Valderrama was observing the solar disk from 12:25 until 13:00 (local time). According to his report, he was looking at a large spot in the eastern region of the southern solar hemisphere. The spot was elongated due to the foreshortening resulting from the proximity to the solar limb. He noted a rounded brightening in the west penumbra that seemed to be the source of a bright band that crossed the active region just south of the umbra.

A short description of the observation was published in the French journal *L'Astronomie* (Valderrama, 1886). The logbook of this observation is far more detailed and complete (see Appendix 1 and its English translation in Appendix 2), and it includes the drawing by Valderrama shown in Figure 1.

Figure 2 shows a photograph of the solar disk from Ogyalla observatory corresponding to the date of observation of the WLF by Valderrama (10 September 1886), available in the web page http://fenyi.solarobs.csfk.mta.hu/en/databases/GPR/ (described by Baranyi, Győri, and Ludmány, 2016). We present the mirror image to maintain the correct orientation. At the western limb, we can see the sunspot described by Valderrama surrounded by a large facular area. In his logbook of solar observations, Valderrama pointed out other examples of limb sunspots surrounded by faculae, but this is the only case where he remarked an unusual brightening associated with the event.

## 3. Discussion

In order to put into context the WLF reported by Valderrama, we analyze the geomagnetic response, the duration of the phenomenon, the exact location of the active region on the solar disk, and the phase of the solar cycle when it happened.

No geomagnetic storms or aurorae were reported in the days following the occurrence of the flare reported by Valderrama (Jones, 1955; Vázquez *et al*. 2014, 2016). Several aspects could explain this fact: i) the energetic flare was not accompanied by a coronal mass ejection (see, for example, Gosling, 1993; Guedes *et al*., 2015), ii) the coronal mass ejection did not reach the Earth (see, for example, Baker *et al*., 2013), or iii) the field orientation and speed of the solar wind were not favorable to induce storms and/or aurorae.

From the time span reported by Valderrama, we can estimate a minimum duration of 30 minutes. There is other historical case for such a long duration WLF. It occurred on 31 March 1938, and the estimated duration was at least 40 minutes (Dobbie, Moss and Thackeray, 1938).

According to the Greenwich Photoheliographic Results (GPR), the sunspot group (number 1923) associated with the WLF observed by Valderrama was located very close to the solar limb (Figure 2). Its heliographic coordinates were latitude 12.7º and longitude 44.2º. The whole area around the sunspot was 190 millionths of the solar disk. Note that Valderrama indicated erroneously the sunspot group was in the southern hemisphere. This error is due to the fact that the sunspot group is located in the southern part of the solar disk, if we divide it into two parts according to the terrestrial E-W direction; see the black line crossing the solar disk in Figure 2 (lower panel). The sunspot group is located in the northern hemisphere if we use the correct solar equator (the red line in the lower panel of Figure 2).

The WLF observed by Valderrama occurred in the descent phase of the solar cycle 12. Historical WLFs do not seem to prefer a particular phase of the cycle to appear. Figure 3 shows the smoothed sunspot number (version 2, Clette *et al*., 2016) from 1850 to 1950. The dates of occurrence of the historical WLFs from Neiding and Cliver, (1983) have been marked using thin vertical lines. The date of Valderrama's WLF has been represented by a thick vertical line. Note that historical WLFs have occurred at the ascent, maximum, and descent phases of the solar cycles. In any case, the WLF observed by Valderrama occurred in a month of relatively low solar activity because the smoothed sunspot number for September 1886 is only 34.2.

## 4. Conclusion

A WLF was observed by Valderrama on 10 September 1886. Chronologically, it is the third solar flare observed in the history of astronomy, after the solar flares observed on 1 September 1859 (Carrington, 1860; Hogdon, 1860; Cliver and Dietrich, 2013), and 13 November 1872 (Secchi, 1872). It was not geoeffective, and occurred in a context of relatively low solar activity. A critical revision of the astronomical literature of the 19th century and early 20$^{th}$ century should increase the list of historic WLFs.

**Disclosure of Potential Conflicts of Interest**

The authors declare that they have no conflicts of interest.

**Acknowledgements**

We thank the IAC librarian, M. Gómez for putting at our disposal the materials used in this study. They were found by J. A. Bonet, and his participation in the early phases of the project is acknowledged. The support from the Junta de Extremadura (Research Group Grants GR15137) and from the Ministerio de Economía y Competitividad of the Spanish Government (AYA2014-57556-P and AYA2013-47742-C04-02-P) is gratefully acknowledged.

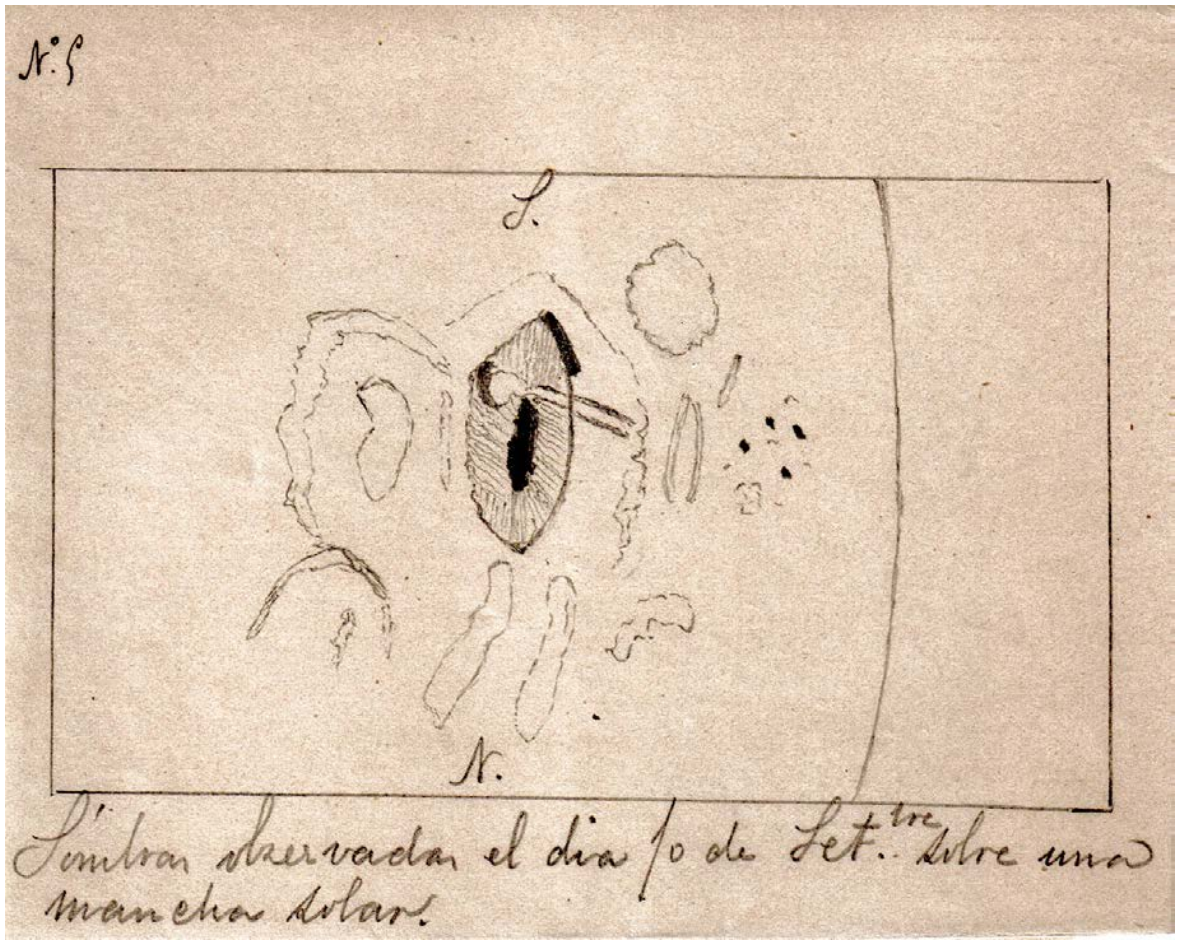

Figure 1. Drawing by Valderrama showing the WLF observed on 10 September 1886. South and north are marked with S and N, respectively, and the nearest solar limb is represented by a curved line. The central sunspot, with its striped penumbra and black umbra, is clear. The WLF is the tadpole-like structure with the head at the center of the penumbra and with the tail towards the limb. The original of this document is preserved at the IAC Library. The translation of the original Spanish caption in the drawing made by Valderrama is: Shadows observed on the day 10 September on a sunspot.

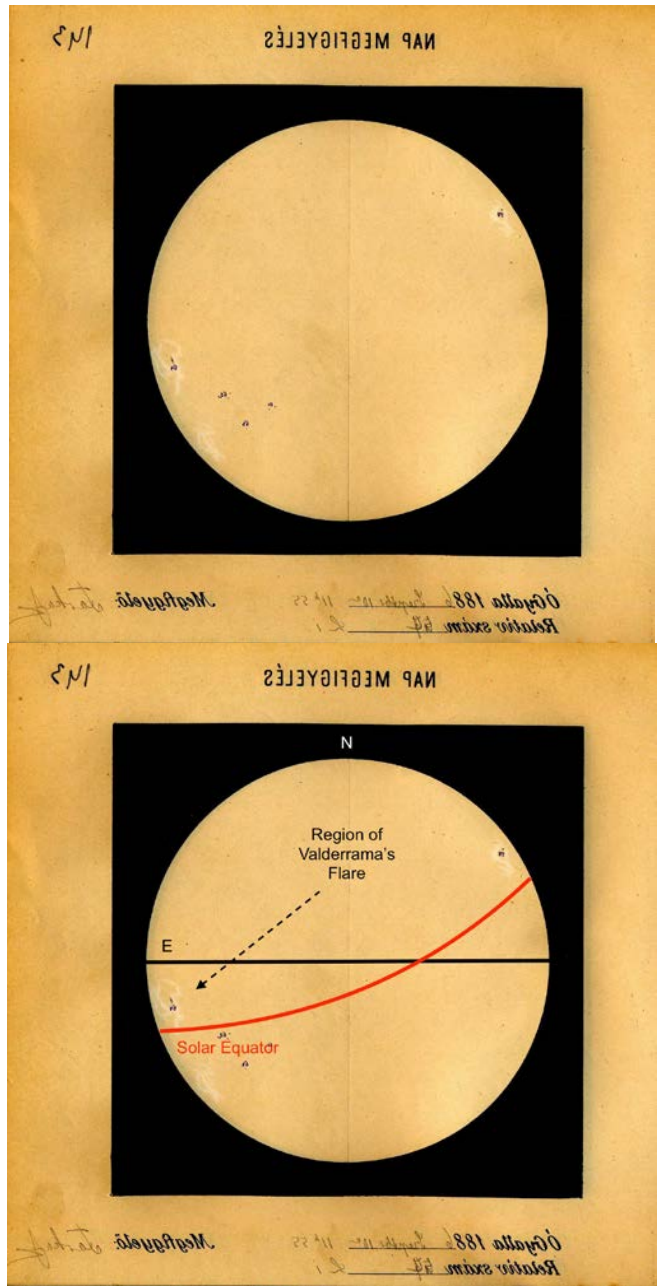

Figure 2. Top panel: photograph from Ogyalla observatory corresponding to the date of Valderrama's WLF, 10 September 1886. Bottom panel, original photograph with the E-W terrestrial line marked in black. Note that the active region with Valderrama's flare was below this line, and so to the south, but still above the solar equator (the red solid line), and so it has positive solar latitude. The image has been downloaded from the web page http://fenyi.solarobs.csfk.mta.hu/en/databases/GPR/ , which is described by Baranyi, Győri, and Ludmány (2016).

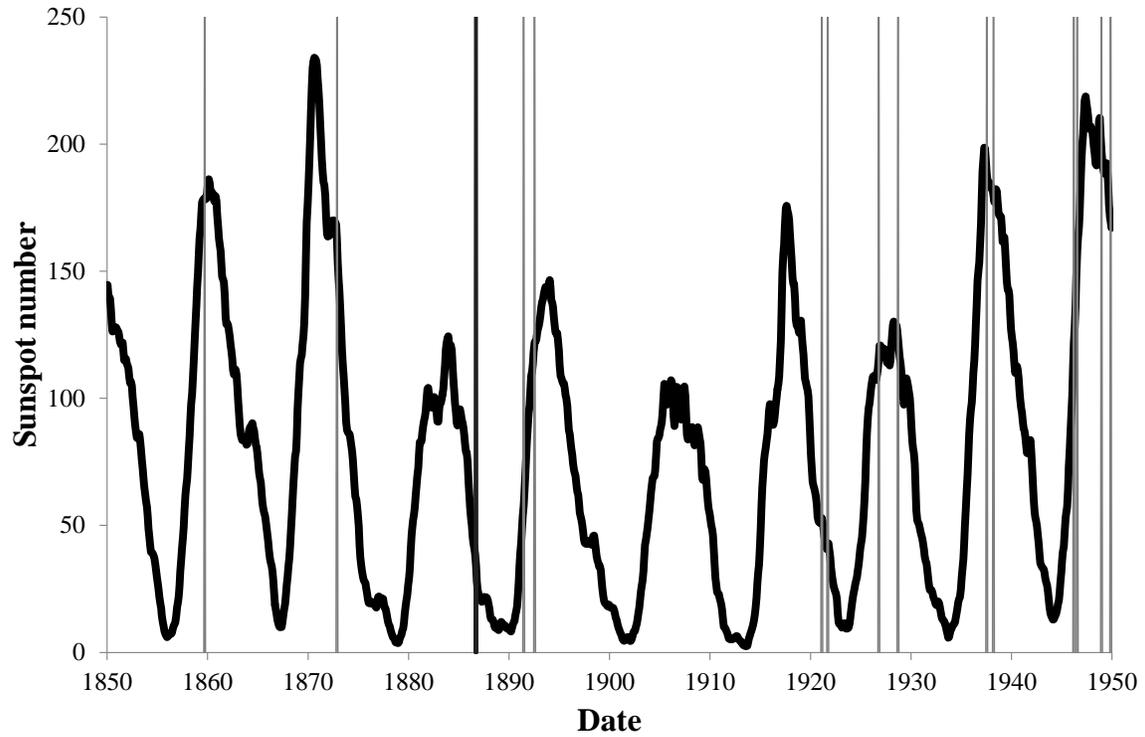

Figure 3. Sunspot number during the period from 1850 to 1950. The thick vertical line indicates the observing date of Valderrama's WLF (10 September 1886). The thin vertical lines point out observing dates of other historical WLFs compiled by Neiding and Cliver (1983).

Matthews, S.A., van Driel-Gesztelyi, L., Hudson, H.S., Nitta, N.V.: 2003, A catalogue of white-light flares observed by Yohkoh. *Astron. Astrophys.* **409**, 1107. DOI: 10.1051/0004-6361:20031187

Neidig, D.F., Cliver, E.W.: 1983, *A catalog of solar white-light flares, including their statistical properties and associated emissions, 1859–1982*, Air Force Geophysics Laboratory Technical Report.

Oliver, J.M.: 1997, *Historia de la astronomía amateur en España*, Equipo Sirius, Madrid.

Rudaux, L.: 1892, La Tache Solaire du 15 Juillet. *L'Astronomie* **11**, 342.

Secchi, A.: 1872, *Compt. Rend. Acad. Sci. Paris* **75**, 1581.

Trouvelot, E.L.: 1891, Phenomene Lumineux Extraordinaire Observe sur le Soleil. *L'Astronomie* **10**, 287.

Valderrama, J.: 1886, Ombres observées sur une tache solaire. *L'Astronomie* **5**, 388.

Vaquero, J.M., Vázquez, M.: 2009, *The Sun Recorded Through History*, Springer, Berlin. DOI: 10.1007/978-0-387-92789-3

Vázquez, M., Vaquero, J.M., Gallego, M.C.: 2014, Long-term Spatial and Temporal Variations of Aurora Borealis Events in the Period 1700–1905. *Solar Phys.* **289**, 1843. DOI: 10.1007/s11207-013-0413-6

Vázquez, M., Vaquero, J.M., Gallego, M.C., Roca Cortés, T., Pallé, P.L.: 2016, Long-Term Trends and Gleissberg Cycles in Aurora Borealis Records (1600‐2015). *Solar Phys*. **291**, 613. doi: 10.1007/s11207-016-0849-6

Wang, H.-M.: 2009, Study of white-light flares observed by Hinode. *Research in Astron. Astrophys*. **9**, 127. DOI: 10.1088/1674-4527/9/2/001
9

**Appendix 1. Original Description by Valderrama in Spanish**

Día 10 de septiembre

Observación del disco solar: 12h 25m á 1h de la tarde

En la región oriental del hemisferio austral se ha formado súbitamente de ayer a hoy una enorme mancha muy bella, está muy alargada por su proximidad al borde... Al observarla con atención he notado sobre ella un fenómeno extraordinario, sobre la penumbra al oeste del núcleo, y casi en contacto con él, se distinguía un objeto muy brillante que producía una sombra perfectamente visible sobre la penumbra de la mancha. Este objeto tenía una forma casi circular, y de su parte oriental partía un largo rayo luminoso que atravesaba la mancha al sur del núcleo, produciendo sombra sobre la penumbra y que iba a perderse sobre la enorme masa de fáculas que rodeaban la extremidad oriental de la mancha. En la extremidad SE de la misma mancha se percibía otra umbra más visible que las precedentes, y que tenía exactamente la misma sombra que la penumbra de la mancha en ese punto. ¿Serían fáculas situadas sobre la penumbra ocultando una parte de la mancha, o bien materias solares flotando sobre la fotosfera, después de haber sido lanzadas por una explosión formidable? Yo creo que el núcleo situado sobre la penumbra de la mancha seguido de un largo rayo luminoso, no es otra cosa, que materias solares flotando encima de la mancha, toda vez que estas se han formado súbitamente en el espacio de unos días. Por el contrario la sombra observada en la extremidad oriental de la mancha, es producida por masas faculares que comienzan a invadirla en ese punto. El dibujo que yo hice (VER DIBUJO), con el más escrupuloso esmero, representa fielmente este fenómeno, uno de los más curiosos que puede ofrecer el astro del día. Al este de la mancha en cuestión se distinguen 4 poros formando un cuadrilátero. Al SO de esta misma mancha se distingue otro grupo formado por 9 poros, otro se distingue al sur sobre un grupo de fáculas y por último otro apenas visible se distingue al SO del anterior. Casi sobre el borde del disco, en la región occidental del hemisferio boreal se columbra aun la gran mancha de los últimos días. Ya no se distingue sino una pequeña parte del núcleo y la mitad de la penumbra.

**Appendix 2: English Translation of the Original Description in Spanish**

The 10 December.

Observations of the solar disk from 12h 25m to 1h in the afternoon.

In the eastern region of the southern hemisphere a huge, beautiful sunspot was formed from yesterday to today, it is elongated due to its proximity to the limb ... by looking at it carefully

I noticed an extraordinary phenomenon on her[1], on the penumbra to the west of the nucleus[2], and almost in contact with it, a very bright object was distinguishable producing a shadow clearly visible on the sunspot penumbra. This object had an almost circular shape, and a light beam came out from its eastern part that crossed the sunspot to the south of the nucleus, producing a shadow on the penumbra that was lost in the large mass of faculae surrounding the eastern extreme of the sunspot. In the SE extreme of the same sunspot another umbra was perceived even more clearly than the preceding ones, which had exactly the same shadow as the sunspot penumbra at that point. Could they be faculae located on the penumbra hiding a part of the sunspot, or rather solar matter floating over the photosphere, after being ejected by a formidable explosion? I believe that the nucleus located on the sunspot penumbra followed by a long light beam is nothing else but solar matter floating on the sunspot, each time that these[3] were formed suddenly in a few days. On the contrary, the shadow observed in the eastern extreme of the sunspot is produced by facular matter that starts to invade it[4] at this point. The drawing I did (see the drawing[5]), with the most scrupulous care, faithfully represents this phenomenon, one of the most curious that the daystar can offer. To the east of the sunspot, 4 pores forming a quadrilateral can be distinguished. To the SW of the same sunspot, another group formed by 9 pores can be distinguished too, another one is distinguished to the south on a group of faculae and, finally, another one barely visible to the SW of the previous one. Almost at the limb of the disk, in the western region of the northern hemisphere, one can discern the last big sunspot of the last days. One can barely distinguish a small part of the nucleus and half of the penumbra.

---

[1] Translators' note: sunspot is feminine in Spanish, and Valderrama refers to the sunspot as her.
[2] Translators' note: umbra.
[3] Translators' note: these structures.
[4] Translators' note: the sunspot.
[5] Translators' note: see Figure 1.